\documentstyle[aps,epsfig,float,twocolumn,prl]{revtex}
\newcommand{\be}{\begin{equation}}
\newcommand{\ee}{\end{equation}}

\begin{document}
\twocolumn[\hsize\textwidth\columnwidth\hsize\csname @twocolumnfalse\endcsname
\draft
\title {Locality in Quantum and Markov Dynamics on Lattices and Networks}
\author{M. B. Hastings}
\address{
T-13, Los Alamos National
Laboratory, Los Alamos, NM 87545, hastings@cnls.lanl.gov 
}
\date{March 28, 2004}
\maketitle
\begin{abstract}
We consider gapped systems governed by either
quantum or Markov dynamics, with
the low-lying states below the gap being approximately degenerate.
For a broad class of dynamics, 
we prove that ground or stationary state
correlation functions can be written as a piece 
decaying exponentially in space plus a
term set by matrix elements
between the low-lying states.  The key to the proof is a local approximation
to the negative energy, or annihilation, part of an operator in a gapped system.
Applications to numerical
simulation of quantum systems and to networks are discussed.

\vskip2mm
\end{abstract}
]

The relation between energy and length scales is central to our understanding of
physics.  We intuitively associate low energies with long wavelengths.  
For many-body systems, at a quantum critical point, where the energy gap 
vanishes, we expect to see long-range
correlations\cite{qcrit}.  Conversely, experience teaches us that a gap in a
quantum system implies a finite correlation length.

Such a result is well known for non-interacting systems; for example,
a defect in a perfect crystal may give rise to localized modes
lying within a band gap.
The basic result shown in \cite{lhd} is that for a large
class of short-range many-body Hamiltonians, this result still holds: if there
is a gap between the ground and first excited states, then the connected correlation 
function of two operators decays
exponentially in space with correlation length $\xi$ bounded by
a characteristic velocity divided by the gap, $\Delta E$.  
We note that this contrasts with the possibility of
having an infinite entanglement length in a gapped system, as is studied in the
field of quantum computation\cite{qcomp}.  

In this Letter, we extend this result to a more general class of systems.  We
consider the case of a quantum system with
some number of almost degenerate low energy states, all
within
energy $\Delta E_{low}$ of the ground
state energy, with the rest of the spectrum having an energy at least
$\Delta E$ above the ground state.  Then, we prove that the 
correlation functions
include an exponentially decaying piece with correlation length
$\xi$, plus a piece which involves matrix elements between the states below the
low-lying states.
The results in this Letter
apply to finite range lattice Hamiltonians, with some technical
conditions required to bound the group velocity on the lattice, as
discussed more below.
This set of systems includes
short-range lattice spin systems, lattice fermion systems, and lattice hard-core
boson systems.
This result provides a general proof of
Kohn's idea of ``nearsightedness"\cite{wk} for this class of systems, and
thus may have important applications in ${\cal O}(N)$ methods for simulating
quantum systems.

Proofs of some of the results used in this Letter can be found in
\cite{lhd}, and additional proofs will be given in \cite{tbp}.  After
giving the basic results, we discuss a wide variety of applications:
to systems with a band structure; to 
classical Markov processes;
and to systems on general networks.

Consider a given connected correlation function: $\langle A B \rangle$, where
$A,B$ are operators and $\langle ... \rangle$ denotes the ground state expectation
value.  
Using a spectral representation, we have $\langle A B \rangle=
\sum_{E_i\leq \Delta E_{low}} A_{0i}B_{i0} + 
\sum_{E_i\geq \Delta E} A_{0i}B_{i0}$, where $i$ represents different intermediate
states and $0$ is the ground state.  Define $(A_{low})_{ij}=A_{ij}$ if both $E_i$ and $E_j$ are less
than or equal to $\Delta E_{low}$, while
$(A_{low})_{ij}=0$ otherwise.  

In this Letter we determine
which correlation functions may be long-range, given the structure of low energy
states.  The basic result is that for a pair of operators
$A,B$ separated by distance $l$, for vanishing $\Delta E_{low}$,
\be
\label{expctd}
\langle A B \rangle = \langle \frac{1}{2}\{A_{low},B_{low}\}\rangle+
{\cal O}(\exp[-l/\xi]).
\ee

A quantum Ising system with a transverse field is a good
example system to apply this result: 
${\cal H}=J\sum_{d(i,j)\leq 1}S^z_iS^z_j+B\sum_i S^y_i$,
where $S^{a}_i$ are the spin operators on site $i$ for $a=x,y,z$ and
where $d(i,j)$ is some metric on the lattice.
In the paramagnetic phase ($B/J$ sufficiently large), the
system has a unique ground state with a gap to the rest of the spectrum and
so Eq.~(\ref{expctd}) implies that all connected correlation functions
decay exponentially.
In the ferromagnetic phase ($B/J$ sufficiently small), the system has
two almost degenerate low energy states, 
and again has a gap to the rest of the spectrum.  
Operators such as $S^z_i$ have long-range correlations in the ferromagnetic 
phase, due to matrix elements of these operators between
the two low-lying states.
Correlation functions of operators which do not couple the low-lying states, 
such as energy-energy correlation functions, are exponentially decaying.  

{\it States Above the Gap---}
Define
\be
A^{\pm}=\frac{1}{2\pi}\int_{-\infty}^{\infty}
{\rm d}t A(t)\frac{1}{\pm it+\epsilon},
\ee
where $A(t)=\exp(i {\cal H}t) A \exp(-i {\cal H} t)$.  Then, $(A^+)_{ij}=
A_{ij} \Theta(E_i-E_j)$, where $\Theta$ is a step function:
$\Theta(x)=1$ for $x>0$ and $\Theta(x)=0$ for $x<0$, while $\Theta(0)=1/2$.
Thus, 
$A^-$ includes only the negative energy (positive
frequency) matrix elements of $A$.  We may define a similar $O^{\pm}$ for
any operator $O$.

Define $A_{high}=A-A_{low}$
Then,
\begin{eqnarray}
\label{can}
\langle A B \rangle=\langle A_{low} B_{low} \rangle + \langle A_{high}^{-} 
B \rangle
\\ \nonumber
=\langle A_{low} B_{low} \rangle + \langle [A_{high}^-,B] \rangle,
\end{eqnarray}
since $\langle B A_{high}^- \rangle=0$.  
Here, we may view
$A^+$ as a creation operator and $A^-$ as an annihilation operator.

The basic idea of the proof in this Letter is that we will find an approximation
to $A^{-}$ 
such that $(1)$: the error involved in the approximation is small and
$(2)$: the commutator of the approximate operator with $B$ is small.  The approximation
is defined by:
\begin{eqnarray}
\label{tdef}
{\tilde A}(t)\equiv A(t) \exp[-(t\Delta E)^2/(2q)], \\
{\tilde A}^{\pm}
=\frac{1}{2\pi}\int {\rm d}t \, 
{\tilde A}(t)\frac{1}{\pm it+\epsilon},
\end{eqnarray}
where $q$ will be chosen later.  
In general, we may define $\tilde O,{\tilde O}^+$ for
any operator $O$.

Then $({\tilde A}^+)_{ij}=A_{ij}\Theta_q(E_i-E_j)$, where
\be
\label{thqd}
\Theta_q(\omega)= 
\int_{0}^{\infty} \frac{{\rm d}\omega'}{2\pi}
(\sqrt{2 \pi q}/\Delta E)
\exp[-q(\omega-\omega')^2/(2\Delta E^2)].
\ee
The function $\Theta_q$ is equal to
an error function; for large $q$ it approximates the step function $\Theta$.

{\it Finite Velocity---}
Since $A$ and $B$ are separated in space, they commute.
Suppose operators $A,B$ are separated in space by a distance $l$, 
meaning that $l$ is
the shortest distance between any two sites $i,j$, such that some operator on site $i$
appears in $A$ and some operator on site $j$ appears in $B$.
Then, one
expects that the commutator of the operators $A(t)$ and $B$ will be small
for some range of times $t$, with $|t|$ less than or equal to $l$ divided by
a characteristic velocity of the system.  

Specifically, we consider
any lattice Hamiltonian ${\cal H}$ which can be written as a sum 
${\cal H}=\sum_i{\cal H}^i$, where $i$ ranges over lattice sites and where we 
require that: $(1)$
the commutator $[{\cal H}^i,O]=0$ for any operator $O$ which acts only on sites $j$
with $d(i,j)>R$, 
where $R$ is the range of the Hamiltonian; and 
$(2)$
the operator norm $||{\cal H}^i||\leq J$,
for all $i$, for some constant $J$.  
Then, it was shown in \cite{fgv,lhd}
that there exists a function $g(t,l)$, which depends on $J$, $R$, and
the lattice structure, such that 
\be
\label{cbnd}
||[A(t),B(0)]||\leq
||A||||B|| \sum_j g(t,l_j),
\ee
where the sum ranges over sites $j$ which
appear in operator $B$, and $l_j=d(j,i)$ is the distance from $j$ to the closest site 
$i$ in the operator $A$.
It was shown that there exists some constant $c_1$ such
that $g(c_1 l,l)$ is exponentially decaying in $l$ for large $l$\cite{lhd}.  
Further $g$ is symmetric in $t$ so that $g(t,l)=g(-t,l)$ and
$g(t,l)\leq (t/t') g(t',l)$ for $t<t'$ and $t,t'>0$.  Thus, $g(t,l)$ is
monotonically increasing in $t$ for $t>0$.  Hence, for $t\leq c_1 l$,
$||[A(t),B(0)]||$ is exponentially decaying in $l$ for large $l$.
We define $v\equiv c_1^{-1}$
as a velocity of the system.
We will define 
\be
\label{cdef}
C(t)=\sum_j g(t,l_j),
\ee
where the sum ranges over sites $j$ which as above.
In a later section, we will discuss the applicability of this result to systems
on general graphs, or {\it networks}.

As a side note, this bound on the operator norm of the commutator may be viewed
as a non-relativistic analogue of the concept of micro-causality in relativistic
field theories\cite{mcc}.  In relativistic theories, operators commute outside
the light-cone, so that if $c$ is the speed of light, then $[A(t),B(0)]=0$ for
$t<l/c$.  In a non-relativistic theory, the commutators do not exactly vanish, but
we are able to find a velocity $v$ such that the commutator is exponentially
small for $t<l/v$.

{\it Commutator---}
To evaluate the commutator in Eq.~(\ref{can}), we use
\begin{eqnarray}
\label{ob}
\langle [A_{high}^-,B] \rangle=\langle [{\tilde A}^-,B] \rangle +
\Bigl( \langle [A_{high}^-,B] \rangle-
\langle [{\tilde A}_{high}^-,B] \rangle\Bigr) \\ \nonumber
-\langle [{\tilde A}_{low}^-,B] \rangle.
\end{eqnarray}
Here, ${\tilde A}_{high}^-$ is defined by starting with $A_{high}$, and
then multiplying by $\exp[-(t\Delta)^2/(2q)]$, following
Eq.~(\ref{tdef}), and finally taking the negative energy part.  Thus,
${\tilde A}_{high}^-+{\tilde A}_{low}^-={\tilde A}^-$.

First we bound the first term on the right-hand side of Eq.~(\ref{ob}).  We have
\begin{eqnarray}
\label{b1}
|\langle [{\tilde A}^-,B] \rangle| 
\\ \nonumber
\leq  
\frac{1}{2\pi}|\int_{|t|<c_1 l} 
{\rm d}t 
\exp[-(t\Delta E)^2/(2q)]
\langle [A(t),B]\rangle \frac{1}{-it+\epsilon}| \\ 
\nonumber
+\frac{1}{2\pi}|\int_{|t|>c_1 l} 
{\rm d}t 
\exp[-(t\Delta E)^2/(2q)]
\langle [A(t),B] \rangle \frac{1}{-it+\epsilon}| \\ 
\nonumber
\leq
\frac{1}{2\pi}||A|| ||B||\Bigl( 2C(c_1 l)+2\frac{\sqrt{2\pi q}}{\Delta E c_1 l}
e^{-(c_1 l \Delta E)^2/(2q)}\Bigr).
\end{eqnarray}
In Eq.~(\ref{b1}), for $|t|< c_1 l$, we have used 
$\exp[-(t\Delta E)^2/(2q)]
\langle [A(t),B \rangle] \rangle \leq
|| [A(t),B \rangle] ||$ and Eq.~(\ref{cbnd}), while for $|t| > c_1 l$ we have
used $\exp[-(t\Delta E)^2/(2q)] \langle [A(t),B \rangle] \rangle \leq
2 \exp[-(t\Delta E)^2/(2q)] ||A|| ||B||$, and then we have performed the
integrals using these bounds.

Next, we consider the second pair of terms on the right-hand of Eq.~(\ref{ob}).
From Eq.~(\ref{thqd}) it follows that, for $|\omega|\geq \Delta E$,
$|\Theta_q(\omega)-\Theta(\omega)|\leq \exp[-q/2]/\sqrt{2\pi q}$.
Thus, if $\Psi_0\rangle$ is the ground state wavefunction,
$|\langle \Psi_0 A^-_{high} -
\langle \Psi_0 \tilde A^-_{high}| \leq ||A||\exp[-q/2]/\sqrt{2\pi q}$, and
also
$|A^-_{high} \Psi_0 \rangle-
\tilde A^-_{high} \Psi_0 \rangle| \leq ||A||\exp[-q/2]/\sqrt{2\pi q}$.  Then,
\be
\label{diffb}
|\langle [A_{high}^-,B] \rangle-
\langle [\tilde A_{high}^-,B] \rangle| \leq 
2 ||A|| ||B|| \frac{\exp[-q/2]}{\sqrt{2 \pi q}}.
\ee
Physically, Eq.~(\ref{diffb})
is a kind of uncertainty relation:
the time integral in Eq.~(\ref{tdef}) extends over a time of order 
$\sqrt{q}\Delta E^{-1}$,
and thus provides an approximation to energies of order
$\Delta E/\sqrt{q}$.

{\it Almost Degenerate Low Energy States---\\}
Suppose $\Delta E_{low}$ is very small compared to $\Delta E$.
Then, for $\omega\leq \Delta E_{low}$,
$\Theta_q(\omega)$ is close to $1/2$:
$|\Theta_q(\omega)-1/2|\leq \sqrt{q/2 \pi}(\omega/\Delta E)$.
Thus, 
$|\langle [{\tilde A}_{low}^-,B] \rangle-\frac{1}{2}
\langle [A_{low},B] \rangle|\leq 
2 ||A_{low}|| ||B_{low}|| \sqrt{q/2 \pi}
(\omega/\Delta E)$. 
Then, from Eqs.~(\ref{can},\ref{ob},\ref{b1},\ref{diffb}) above, it follows
that
\begin{eqnarray}
\label{fb1}
|\langle A B \rangle-\langle A_{low} B_{low} \rangle-\frac{1}{2}
\langle [A_{low},B_{low}] \rangle |\\
\nonumber
\leq
2||A|| ||B||\Bigl(\frac{C(c_1 l)}{2\pi}+
\frac{\sqrt{q}}{2 \pi \Delta E c_1 l}
\exp[-(c_1 l \Delta E)^2/(2q)]\\ \nonumber
+\frac{\exp[-q/2]}{\sqrt{2\pi q}}
\Bigr)
+2 ||A_{low}|| ||B_{low}|| 
\sqrt{q/2 \pi}
(\Delta E_{low}/\Delta E).
\end{eqnarray}

Picking $q=c_1 l \Delta E$, we have
\begin{eqnarray}
\label{ed}
|\langle A B \rangle-\langle \frac{1}{2}\{A_{low},B_{low}\}\rangle| \\ \nonumber
\leq
||A|| ||B||\Bigl[2 C(c_1 l)/(2\pi)+
\frac{4}{\sqrt{2\pi q}}
e^{-c_1 l \Delta E/2} \Bigr]\\ \nonumber
+2 ||A_{low}|| ||B_{low}|| 
\sqrt{c_1 l \Delta E/2 \pi}
(\Delta E_{low}/\Delta E).
\end{eqnarray}
The first term on the right-hand side of Eq.~(\ref{ed}) is exponentially
decaying in $l$ with some correlation length $\xi_C$.  The terms in
parenthesis are exponentially decaying with correlation length
$2/(c_1 \Delta E)$.  Thus, we define the correlation length $\xi$ to
be the minimum of $\xi_C$ and $2/(c_1 \Delta E)$.  For $\Delta E_{low}$ taken to
be zero, we have Eq.~(\ref{expctd}).

{\it Low Energy Commutator---}
We can also bound the expectation value
$\langle [A_{low},B_{low}] \rangle$.  Define $\tilde A^0=
\Delta E/\sqrt{2\pi q}
\int_{-\infty}^{\infty} {\rm d}t \, \tilde A(t)$.  Then, 
we can bound the commutator: $|\langle [\tilde A^0,B] \rangle|
\leq 
\Delta E/\sqrt{2\pi q}
\Bigl(\int_{|t|<c_1 l}{\rm dt}\, C(t)+
|\int_{|t|>c_1 l}{\rm d}t\, \langle[\tilde A(t),B]|\rangle
\leq 
||A|| ||B|| \Bigl(\sqrt{c_1 l \Delta E/2\pi} C(c_1 l)+
2 e^{-c_1 l \Delta E/2}\Bigr)$.

Next, consider the difference
$\langle [\tilde A^0,B] \rangle-\langle [A_{low},B_{low}]
\rangle$.  This equals
$(\langle [\tilde A_{low}^0,B \rangle - \langle [A_{low},B_{low}]\rangle)
+\langle [\tilde A_{high}^0,B]\rangle$.  Explicit computation with a
spectral representation gives
$|\langle [\tilde A_{low}^0,B \rangle - \langle [A_{low},B_{low}]\rangle|\leq
2 ||A_{low}|| ||B_{low}|| 
(e^{-q(\Delta E_{low}/\Delta E)^2/2}-1)$, while
$|\langle [\tilde A_{high}^0,B]\rangle|\leq 2||A|| ||B|| e^{-c_1 l \Delta E/2}$.
Combining these bounds with the bound on $|\langle [\tilde A^0,B] \rangle|$ 
gives
\begin{eqnarray}
\label{flb}
|\langle [A_{low},B_{low}] \rangle|
\leq \sqrt{c_1 l \Delta E/2\pi} ||A|| ||B|| C(c_1 l)\\ \nonumber+ 
4||A|| ||B|| e^{-c_1 l \Delta E/2}
\\ \nonumber
+2 ||A_{low}|| ||B_{low}|| 
(e^{-q(\Delta E_{low}/\Delta E)^2/2}-1).
\end{eqnarray}
Thus, for $\Delta E_{low}=0$, we find that
$|\langle [A_{low},B_{low}] \rangle|$ is exponentially decaying in $l$.

{\it Operators at Different Times---}
We finally extend the result Eq.~(\ref{ed})
to correlation functions
$\langle A(-i\tau) B(0) \rangle$, with $\tau$ real and $\tau>0$.
Define 
\be
\tilde A^{\pm}(\pm i\tau)=\frac{1}{2\pi}\int {\rm d}t\, \tilde A(t) 
\frac{1}{\pm it+\tau}.
\ee
by $-it+\tau$.  In this case, we find that Eq.~(\ref{b1}) still
holds as a bound for $|\langle [\tilde A^-(-i\tau),B]\rangle|$.
One may also show that,
for $\tau\leq q/\Delta E$,
$|\langle[\tilde A_{high}^-(-i\tau),B]\rangle-\exp[+\tau^2\Delta E^2/(2 q)]
\langle [A_{high}^-(-i\tau),B]\rangle|\leq 
2 ||A|| ||B|| e^{-q/2}$.

With the given $q=c_1l\Delta E$
the above bounds show that for $\tau\leq c_1 l$,
$|\langle A(-i\tau)
B(0) \rangle
-\langle A_{low}(-i\tau) B_{low}(0) \rangle|\leq$
\begin{eqnarray}
\label{tsep}
e^{-\tau^2\Delta E/(2 c_1 l)}
[\frac{1}{2\pi}
2 ||A|| ||B|| C(c_1 l) + \\ \nonumber
(2+\frac{2}{\sqrt{2 \pi q}}) ||A|| ||B||
e^{-c_1 \Delta E l/2} +
|\langle [\tilde A_{low}(-i\tau),B] \rangle|.
\end{eqnarray}

To bound the last term in the above Equation, we use
$||A_{low}(t)-A_{low}(0)||
\leq t\Delta E_{low} ||A_{low}||$.
Define $z=(2\pi)^{-1}\int {\rm d}t \, \exp[-(t\Delta E)^2/(2q)]/(-it+\tau)$.
Then,
$|\langle [\tilde A_{low}^-(-i\tau),B] \rangle-
z \langle [A_{low},B_{low}] \rangle|\leq 
\pi^{-1}\int {\rm d}t \, t 
\Delta E_{low} ||A_{low}|| |||B_{low}|| \exp[-(t\Delta E)^2/(2q)]/(-it+\tau)\leq
\sqrt{2q/\pi}||A_{low}|| ||B_{low}|| (\Delta E_{low}/\Delta E)$.  
Thus, for $\Delta E_{low}=0$, 
$|\langle [\tilde A_{low}(-i\tau),B] \rangle|$
is exponentially decaying in $l$, following
Eq.~(\ref{flb}).

{\it Band Structure---}
Another interesting application of the above techniques is to problems
with a band structure.  
Note that all of the results above can be generalized
to fermionic operators $A,B$ by interchanging commutators and anti-commutators
throughout.
Suppose we have a free fermionic
theory, with a spectrum which has two bands with a band gap
of width $2\Delta E$.  
Then, we can shift the zero of energy so that the
spectrum has some set of states with energy at most $-\Delta E$ and
another with energy at least $\Delta E$.  Then, if $A=\psi^{\dagger}_i$
is the fermionic creation
operator at some point $i$, we can define an operator 
$\tilde A^-$ which approximately
projects $\psi^{\dagger}_i$ onto the lower band.
At the same time, $\tilde A^{-}$ will
be exponentially localized around point $i$, so that 
$||\{\tilde A^{-},O\}||=
{\cal O}(\exp[-(c_1 l \Delta E)^2/(2q)])+{\cal O}(C(c_1 l))$, 
if the fermionic operator
$O$ acts only on a site $j$ with
$d(x,j)=l$.  This projection technique may be a useful way to compute the 
density matrix in these systems\cite{amn}: if the chemical potential is
such that all states are filled up to zero energy, then
$\rho(i,j)\equiv\langle \psi^{\dagger}_i \psi_j \rangle=
\langle \{A^-,\psi_j\} \rangle \approx 
\langle \{ {\tilde A}^-,\psi_j \} \rangle.$

{\it Markov Processes---}
The above locality results can also be carried over to systems which
obey
continuous time dynamics, following \cite{lhd}, where we have
a transition matrix $T_{ij}$ and a
probability $p_i$ of being in state $i$, so that $\partial_t p_i=\sum_j
T_{ij} p_j$.  By conservation of total probability, we have $\sum_{i}
T_{ij}=0$, guaranteeing that $T$ has at least one zero eigenvalue.  Let
the stationary state with zero eigenvalue have right eigenvector $p_i^0$;
let $I_i$ be the vector with $I_i=1$ for all $i$; this is a left eigenvector
of $T$ with eigenvalue $0$.  

Suppose all eigenvalues of $T$ are real (this includes all systems
for which the stationary state obeys detailed balance).
All eigenvalues of $T$ are non-positive.  Then, assume that there are 
some number of eigenvalues $\lambda_i$ of $T$ with $0\geq \lambda_i\geq
-\Delta_{low}$, while all other eigenvalues $\lambda_i$ have
$\lambda_i\leq -\Delta$, with $\Delta>\Delta_{low}$.  
For each quantity to be measured, $A,B,...$,
define $\langle A \rangle=\sum_i A_i p_i^0$.  We can introduce for
each quantity a diagonal matrix given by $\hat A_{ii}=A_i$, and
$\hat A_{ij}=0$ for $i\neq j$.  Then, $\langle A(t) B(0) \rangle=
I^{\dagger} \exp[-T t] \hat A \exp[T t] \hat B p^0\equiv
I^{\dagger} \hat A(t) B p^0$.  We have
left off the indices on the vector $I,p$ and on the matrices $\hat A,\hat B,
\exp[\pm T t]$; the product is evaluated following the usual rules of
matrix multiplication.  
We can continue these definitions
of $A(t)$ to {\it imaginary time}, and define  $\tilde A(it)=\hat A(it)
\exp[-(t\Delta)^2/(2q)]$ while $\tilde A^+=\frac{1}{2\pi}\int {\rm dt}\,
\tilde A(it)(\mp it+\epsilon)^{-1}$.

Assume that $T$ can be written as a sum of matrices $T^i$, with finite
interaction range $R$ and bound $||T^i||\leq J$ for all $i$.
With these preliminaries, all of the above manipulations can be
carried out for Markov processes.  We find, in particular, that
$\langle A B \rangle=\langle (1/2)\{A_{low},B_{low}\}\rangle+
{\cal O}(\exp[-l/\xi])$.

{\it Networks---}
Recently, systems on general graphs or {\it networks} have been much
studied\cite{net}.  Example systems include the random graph and the
small-world network\cite{sw}.  Locality is an important question in these
systems, for community detection, for example\cite{cdet,nrg}.  
In a small-world, the presence of long-range jumps can completely
destroy locality above some length scale so that all critical behavior
is mean-field\cite{mfa}.

The study of quantum systems on networks is a bit unphysical, but Markov
processes are extremely natural to study on networks.  A good example
is the contact process for epidemic spreading\cite{cp}.

Following \cite{lhd}, the bound Eq.~(\ref{cbnd}) holds for any graph
for which all sites have a bounded number of neighbors (counting
a neighbor as any other site
within distance $R$) and for
which $|T^i|\leq J$ for all sites $i$.  This includes a large class of
interesting networks, such as the small-world network.  However, another large
class of interesting networks, the scale-free networks, have an unbounded
coordination number.
In many of these cases, ${\cal H}$ (or $T$) can be written as a sum of
operators lying on bonds which have bounded operator norm.
Then, we can add to the graph a
set of additional sites which lie on these bonds between sites on
the original network, calling one of new these sites $(i,j)$ if it lies on the 
bond
between $i$ and $j$.  Then, we can let ${\cal H}=\sum_{(i,j)} {\cal H}^{(i,j)}$,
with $||{\cal H}^{(i,j)}|| \leq J$. 

However, we still must deal with the unbounded number of neighbors.  The
range of one of these bond Hamiltonians is such that $(i,j)$ is within range
$R$ of $(i,k)$ for all $j,k$ which neighbor $i$, so that we cannot bound the
number of neighbors within range of a given bond $(i,j)$.  However, if
the given scale-free networks is {\it loopless}, inspection of
the proof in \cite{lhd} shows that Eq.~(\ref{cbnd}) still holds.
Thus, we claim that for networks with either bounded coordination number
and bounded ${\cal H}^i$ or loopless networks with bounded ${\cal H}^{ij}$,
the result Eq.~(\ref{expctd}) is valid.

As an application, consider a contact process in the endemic phase on
a small-world network.  The Markov process governing the dynamics
has a gapped transition matrix away from the critical point.  Then,
a correlation function of the number of infected individuals on a site $i$
(this number is zero or one)
with the number on a site $j$ decays exponentially as  $\exp[-d(i,j)/\xi]$ where
$d(i,j)$ is the shortest path distance (including long-range links) from $i$
to $j$.

{\it Discussion---}
The result we have shown is expected on physical grounds.  In a gapped
system, the correlation function can be written as (\ref{expctd}): an 
exponentially decaying piece, plus a piece determined by matrix elements
between the low lying states.  In the quantum Ising system mentioned
in the introduction, the two low-lying states are symmetric and anti-symmetric
combinations of spin-up and spin-down states.  Acting on the system with
a local operator, such as $S^z_i$, changes between these two states, and
so one may have long-range spin-spin correlation functions; these
correlation functions converge exponentially to some constant at large
separations.
In other cases, such as a spin liquid or fractional quantum Hall system
with a topological degeneracy and a gap to the rest of the 
spectrum\cite{wen}, the matrix elements between the low-lying states
are small for all local operators, so all correlation functions decay
exponentially.

{\it Acknowledgements---}
This work was supported by DOE contract W-7405-ENG-36.

\end{document}